\documentclass[useAMS,usenatbib,article]{mn2e}

\usepackage{longtable}
\usepackage[]{graphicx}
\usepackage{amsmath}

\newcommand\aj{{AJ}}%
\newcommand\araa{{ARA\&A}}%
\newcommand\apj{{ApJ}}%
\newcommand\apjl{{ApJ}}%
\newcommand\apjs{{ApJS}}%
\newcommand\aap{{A\&A}}%
\newcommand\mnras{{MNRAS}}%
\newcommand\pasa{{PASA}}%

\title[The $M_{\rm BH}-n$ relations]{The supermassive black hole mass - S\'ersic index relations for bulges and elliptical galaxies}

\author[G. Savorgnan et al.]
{\parbox{\textwidth}{
G. Savorgnan$^{1}$\thanks{E-mail: \texttt{gsavorgn@astro.swin.edu.au}},
A. W. Graham$^{1}$, 
A. Marconi$^{2}$, 
E. Sani$^{3}$, 
L. K. Hunt$^{3}$, 
M. Vika$^{4}$,
and S. P. Driver$^{5,6}$}\vspace{0.4cm}\\
\parbox{\textwidth}{
$^{1}$Centre for Astrophysics and Supercomputing, Swinburne University of Technology, Hawthorn, Victoria 3122, Australia.\\
$^{2}$Dipartimento di Fisica e Astronomia, Universit\`a di Firenze, via G. Sansone 1, I-50019 Sesto Fiorentino, Firenze, Italy.\\
$^{3}$INAF - Osservatorio Astrofisico di Arcetri, Largo E. Fermi 5, I-50125 Firenze, Italy.\\
$^{4}$Carnegie Mellon University in Qatar, Education City, PO Box 24866, Doha, Qatar.\\
$^{5}$ICRAR - The University of Western Australia, 35 Stirling Highway, Crawley, WA 6009, Australia.\\
$^{6}$SUPA - School of Physics \& Astronomy, University of St Andrews, North Haugh, St Andrews, KY16 9SS, UK.\\}}

\pagerange{\pageref{firstpage}--\pageref{lastpage}} \pubyear{2013}

\begin{document}

\maketitle

\label{firstpage}

\begin{abstract}
Scaling relations between supermassive black hole mass, $M_{\rm BH}$, 
and host galaxy properties are a 
powerful instrument for studying their coevolution. 
A complete picture involving \emph{all} of the black hole scaling relations, 
in which each relation is consistent with the others,
is necessary to fully understand the black hole-galaxy connection.
The relation between $M_{\rm BH}$ and the central light concentration of the surrounding bulge, 
quantified by the S\'ersic index $n$, may be one of the simplest and strongest such relations,
requiring only uncalibrated galaxy images.
We have conducted a census of literature S\'ersic index measurements for a sample of 54 local
galaxies with directly measured $M_{\rm BH}$ values.
We find a clear $M_{\rm BH} - n$ relation, 
despite an appreciable level of scatter due to the heterogeneity of the data. 
Given the current $M_{\rm BH} - L_{\rm sph}$ and the $L_{\rm sph} - n$ relations,
we have additionally derived the \emph{expected} $M_{\rm BH} - n$ relations, 
which are marginally consistent at the $2\sigma$ level with the \emph{observed} relations.
Elliptical galaxies and the bulges of disc galaxies are each expected to follow 
two distinct \emph{bent} $M_{\rm BH} - n$ relations
due to the S\'ersic/core-S\'ersic divide.
For the same central light concentration, we predict that $M_{\rm BH}$ in the S\'ersic bulges 
of disc galaxies are an order magnitude higher than in S\'ersic elliptical galaxies
if they follow the same $M_{\rm BH} - L_{\rm sph}$ relation.
\end{abstract}


\section{Introduction}
Observations over the past decade have suggested a strong connection between supermassive black holes
(SMBHs) and their host galaxies, or rather spheroids, in spite of the huge difference between their respective sizes. 
While it is clear that the stories of these two objects -- the black hole and the galaxy -- are tightly interwoven, 
the origin and nature of their link are still a subject of debate. 
The scaling relations between the SMBH mass, $M_{\rm BH}$, and the host spheroid properties
make the study of black hole growth an indispensable ingredient
to understand the more general framework of galaxy formation and evolution.
Beyond the well known relation with the velocity dispersion $\sigma$ 
\citep{fer2000,geb2000,gra2011b}, the masses of SMBHs have been shown to correlate with a wide series
of properties belonging to the spheroidal component of the host galaxy, such as the spheroid luminosity
\citep{kor1995,mcl2002,mar2003,gra2013} and stellar mass \citep{lao2001,sgs2013}, the spheroid dynamical mass 
\citep{mag1998,mar2003,har2004,gra2012} and the central stellar concentration of the spheroid
\citep{gra2001b}. 
The connection between bulge mass and disc galaxy morphological type
means that the pitch angle of a disc galaxy's spiral arms is also
related to the black hole mass (e.g. \citealt{dav2013,ber2013}, and references therein).
The old $M_{\rm BH} \propto \sigma^4$ and $M_{\rm BH} \propto L^{1.4}$
relations were actually inconsistent with each other (e.g. \citealt{lau2007}), 
and inconsistent with the curved $M_{\rm BH} - n$ relation \citep{gra2007} given the existence of a linear $L - n$ 
relation (see Section \ref{sec:predmbhn}).
The first of these inconsistencies was addressed in \citeauthor{gra2012} (\citeyear{gra2012}; see also Section 6
of \citealt{gra2008}), and we tackle the second here.
The astrophysical interest in all of these empirical relations 
resides partly in the fact that they must \emph{all} be taken into account by any complete theory or model 
describing the coevolution of galaxies and SMBHs, and also in their employment to predict 
the masses of SMBHs in other galaxies.

A decade ago \citet{gra2001b} presented evidence for a strong correlation between the stellar light concentration
$C_{\rm r_e}(1/3)$ of spheroids and their SMBH mass, showing that more centrally concentrated spheroids have more massive
black holes. \citet{gra2007} re-investigated the same relation, directly using the \citet{ser1963,ser1968} index $n$ as a measure
of the radial concentration of the stars. 
In addition to a log-linear relation, \citet{gra2007} fit a log-quadratic regression,
finding that the $M_{\rm BH} - n$ relation changed slopes at the low- and high-mass end, and
had a level of scatter equivalent to the $M_{\rm BH}-\sigma$ relation 
at that time ($\sim 0.3~\rm dex$).
The advantages of using the $M_{\rm BH} - n$ relation to predict the mass of SMBHs are
several: as noted by \citet{gra2007}, the measurement of $n$ requires only images (even photometrically uncalibrated);
is not heavily affected by possible kinematic substructure at the center of a galaxy, nor by rotational velocity
or the vertical velocity dispersion of an underlying disc, nor by aperture corrections; it is cheap to acquire in terms of telescope
time; does not depend on galaxy distances. 

\citet{pas2013} have recently pointed out that the recent deep, wide-field photometric surveys of galaxies
-- e.g. the Sloan Digital Sky Survey (SDSS, \citealt{yor2000}) and the Galaxy And Mass Assembly (GAMA, \citealt{dri2011}) --
are providing us with large statistically useful samples of galaxies whose major morphological components can be resolved 
out to $z \simeq 0.1$. 
Furthermore, automatic image analysis routines, 
such as GIM2D \citep{sim2002}, GALFIT \citep{pen2002,pen2010}, BUDDA \citep{gad2008} 
and GALPHAT (Yoon, Weinberg \& Katz, in preparation),
can be used to model the surface brightness
distribution of the stellar components of these galaxies (e.g. \citealt{all2006,sim2011,kel2012}). 
A bulge/disc decomposition, 
along with adequate corrections to account for dust and inclination effects as provided by \citet{pas2013}, 
can provide the S\'ersic index of the 
spheroid component of both elliptical and disc galaxies.
This can then be used to predict black hole masses in large
samples of galaxies
to derive the local black hole mass function (e.g. \citealt{gra2007c}) and space density (\citealt{gra2007b}, and references therein), 
if a well calibrated $M_{\rm BH} - n$ relation exists.
However, in the past two years \citet{san2011}, \citet{vik2012} and \citet{bei2012} have failed to recover a strong 
$M_{\rm BH} - n$ relation.

Due to the existence of the luminosity-$n$ relation (e.g. \citealt{you1994,jer2000,gra2013rev}, and references therein)
and the $M_{\rm BH} - $luminosity relation (e.g. \citealt{mag1998}),
an $M_{\rm BH} - n$ relation must exist\footnote{It is not yet established which are the
primary or secondary relations.}.
It is important to investigate why the $M_{\rm BH} - n$ relation may not have been recovered in the above studies.
It is also important to know how it fits in with, and is consistent with, the other scaling relations.
Not only does a proper and complete understanding of the SMBH-galaxy connection require this,
but the central concentration of stars, reflecting the inner gradient of the gravitational potential, 
should be intimately related to the black hole mass.
A well determined $M_{\rm BH} - n$ relation may also provide an easy and accurate means 
to predict black hole masses in other galaxies.
Eventually, semi-analytic models of galaxy formation and simulations should include in their recipes 
\emph{all} of the black hole mass scaling relations. 

In this work we present a census of literature S\'ersic index measurements 
for local galaxies with directly measured supermassive black hole mass. 
We re-investigate and recover the $M_{\rm BH} - n$ relation using the combined data from four past independent works.
In Section \ref{sec:data} we describe our galaxy sample, and in
Section \ref{sec:res} we present the $M_{\rm BH} - n$ scaling relation 
which is then discussed and compared with predictions in Section \ref{sec:predmbhn}. 
Finally, we summarize our analysis in Section \ref{sec:concl}.

\section{Data}
\label{sec:data}

\subsection{SMBH masses}
\label{sec:mbhdata}
Our SMBH galaxy sample comes from \citet{gra2013}, who have built
a catalog of 80 galaxies with supermassive black hole masses obtained from direct maser, stellar or gas 
kinematic measurements. 
Black hole masses for our final sample are listed in Table \ref{tab:data},
along with their total galaxy B-band absolute magnitudes, $M_{\rm B_{\rm T}}$,
taken from the \emph{Third Reference Catalogue of Bright Galaxies} (\citealt{rc3}, hereafter RC3)
and also their morphological classification. 
The final sample consists of those galaxies for which S\'ersic indices have been reported by at least
one of the four studies mentioned below.

\subsection{Collecting S\'ersic indices}
\label{sec:ndata}
The radial light distribution of spheroidal systems (such as elliptical galaxies or the bulges of lenticular and spiral galaxies)
is well described by the \citet{ser1963,ser1968} $R^{1/n}$ model that parametrizes the intensity $I$
as a function of the projected galactic radius $R$ such that
\begin{equation*}
I(R) = I_{\rm e} \exp \biggl \{ -b_{\rm n} \biggl [ \Bigl (\frac{R}{R_{\rm e}} \Bigr )^{1/n} -1 \biggr ] \biggr \} 
\end{equation*}
(\citealt{cao1993,and1995,gra2005}, and references therein).
The quantity $I_{\rm e}$ is the intensity at the effective radius $R_{\rm e}$ that encloses half of the 
total light from the model, and $b_{\rm n}$ is a constant defined in terms of 
the S\'ersic index $n$, which is the parameter that measures the curvature of the radial light profile.

We obtained S\'ersic index measurements for our SMBH sample from the following four independent works.

i) \citet[][hereafter GD07]{gra2007} fit the radial light profiles from a sample of 27 elliptical and disc galaxies
with SMBH masses derived from resolved dynamical studies.
The light profiles they used were predominantly from \citet{gra2001b}, who
searched the various public archives for high-quality R-band images and fit ellipses to the isophotes
with the IRAF task \texttt{ellipse}, 
allowing the position angle and ellipticity to vary with radius\footnote{A discussion of the original galaxy
light profiles can be found in \citet{erw2004} and \citet{tru2004}.}. 
The resulting light profiles were then fit by GD07 with a seeing-convolved S\'ersic $R^{1/n}$ model for elliptical galaxies,
and with a combined (seeing-convolved) exponential disc and $R^{1/n}$ bulge for the disc galaxies,
using the subroutine UNCMND from \citet{kah1989}. 
The inner couple of arcseconds of the profiles was in some instances excluded from the fit
due to the potential presence of partially depleted cores or AGNs, that would produce 
a biasing central deficit or excess of light relative to the inward extrapolation of their outer S\'ersic profile.

ii) \citet[][hereafter V12]{vik2012} investigated the $M_{\rm BH} - n$ and the $M_{\rm BH} - L$
relations. They performed two dimensional (2D) profiling with GALFIT3 on near-IR images (from the UKIDSS-LASS survey) 
of a sample of 25 galaxies. 
V12 fit the light distribution using a S\'ersic function
for the elliptical galaxies, the bulges and the bars of lenticular/spiral
galaxies, and an exponential function for the disc components. 
In the case of core-S\'ersic galaxies with partially depleted cores, they implemented a mask for the inner region.
Bright nuclei were additionally modelled as point sources using the PSF.
A relation between SMBH mass and the S\'ersic index was not found by V12.
They noticed that the S\'ersic index can vary significantly from study to study
and they suggested that such mismatch may be due to the different weighting 
of pixels during the fit that each study used and/or to a wavelength bias.
The signal-to-noise weighted fitting routines, such as GALFIT, can be highly sensitive to central dust obscuration,
unaccounted for central excesses and deficits of light relative to the fitted model, 
and especially errors in the adopted PSF.

\begin{figure}
\begin{center}
\includegraphics[width=\columnwidth, trim=10mm 10mm 10mm 10mm]{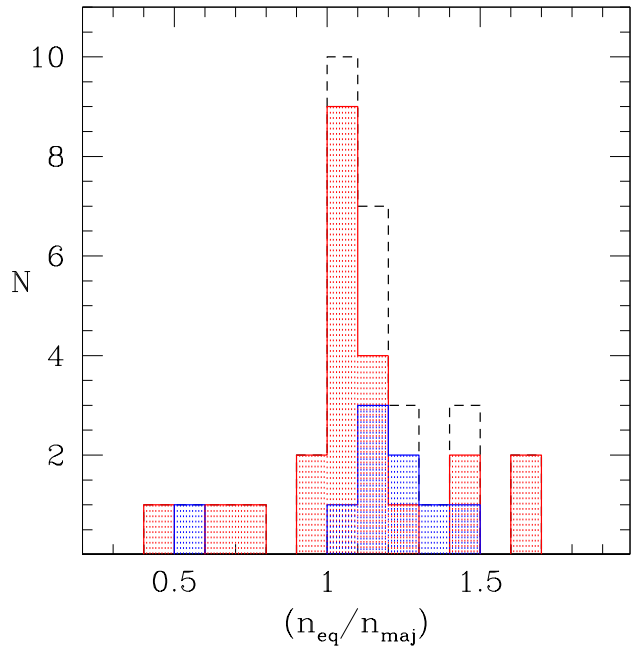}
\end{center}
\caption{Distribution of the ratio between the ``equivalent'' S\'ersic index $n_{\rm eq}$ and that
measured along the major-axis $n_{\rm maj}$. Data are taken from \citet{cao1993}.
The red histogram is for elliptical galaxies, while the blue is for lenticular galaxies 
and the black dashed line represents the whole sample.}
\label{fig:caon}
\end{figure}

\begin{figure}
\begin{center}
\includegraphics[width=\columnwidth, trim=0mm 10mm 10mm 10mm]{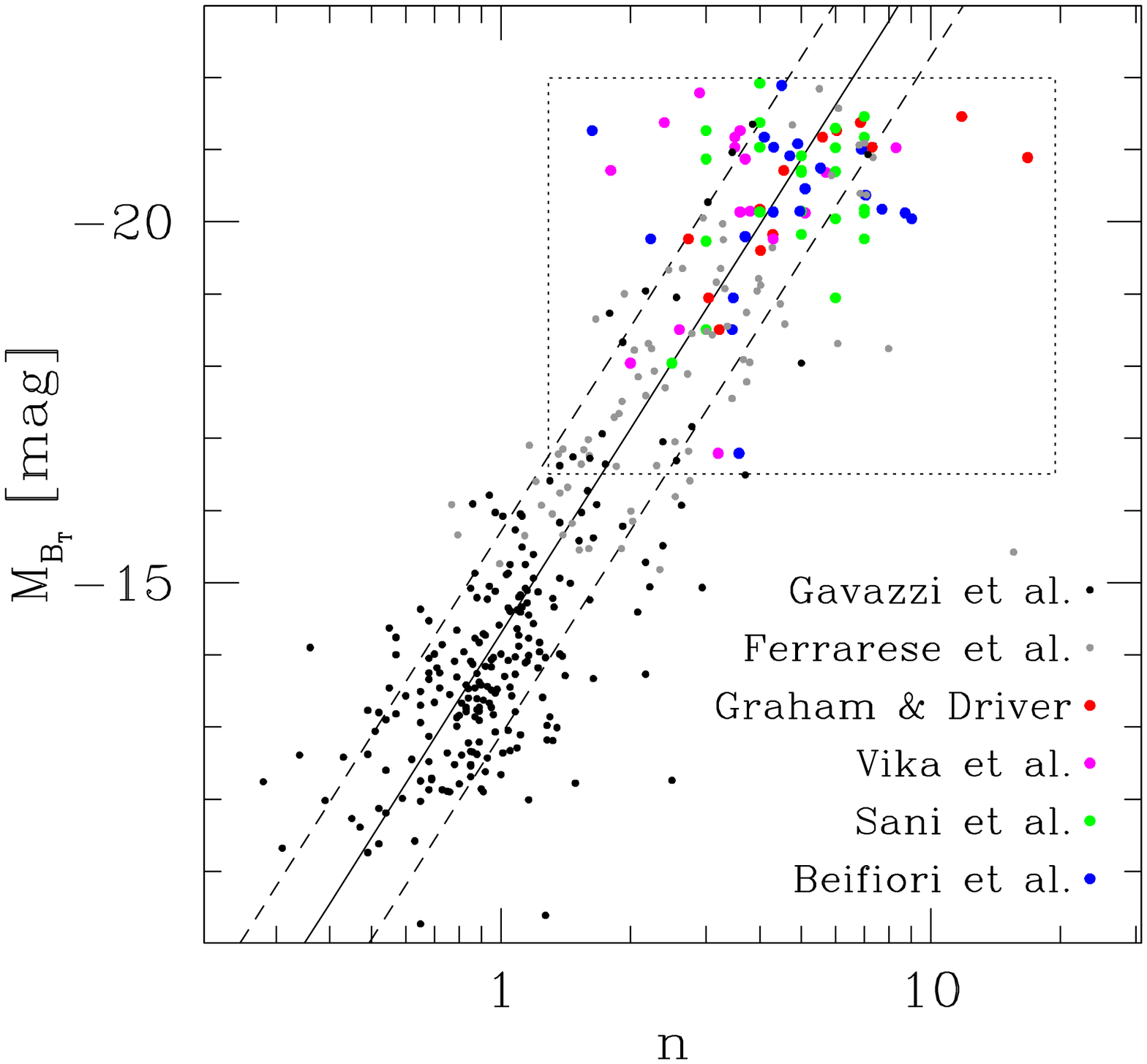}
\end{center}
\caption{Absolute B-band magnitude vs S\'ersic index 
of elliptical galaxies. 
Black points are measurements from \citet{gav2005}; grey points are from \citet{fer2006};
red points are from GD07; pink points are from V12; green points are
from S11; blue points are from B12. 
The black points from \citet{gav2005} and the grey points from
\citet{fer2006} have been plotted just for illustrative purposes, but they will be ignored 
in the following analysis because they are not from a black hole sample.
Each galaxy can have
more than one S\'ersic measurement and hence may be represented more than once along the 
horizontal axis
(with different colours). The black solid line shows the 
elliptical galaxy $M_{\rm B_{\rm T}} - n$ relation from \citet{gra2003a}, 
while the dashed lines are a rough ``by eye'' estimate of the scatter from their diagram. 
The dotted box marks the region that is shown in Figure \ref{fig:bmagnzoom}.}
\label{fig:bmagn}
\end{figure}

iii) From their GALFIT3-derived 2D bulge-disc decompositions of \emph{Spitzer}/IRAC $3.6~\mu\rm  m$ images of 57 galaxies, 
\citet[][hereafter S11]{san2011} investigated the scaling relations between SMBH mass and several other parameters 
of the host spheroids. The image decomposition was performed with a S\'ersic model 
for the elliptical galaxies and with a S\'ersic model plus an exponential model for the lenticular and spiral galaxies.
A Gaussian component and a nuclear point source were added in the presence of a bar or an AGN, respectively.
In an attempt to restrict the degeneracy between the effective radius and the S\'ersic index, 
following \citet{hun2004},
S11 performed 2D fitting by
fixing the S\'ersic index to a set of constant values in the range
between $n=1$ and $n=7$. 
They found tight correlations between the SMBH mass and the bulge luminosity and dynamical mass.
However, the relation
between the SMBH mass and the effective radius had a high intrinsic dispersion and no correlation with the S\'ersic index 
was found.

iv) \citet[][hereafter B12]{bei2012} analyzed Sloan Digital Sky Survey $i$-band images and extracted photometric and structural
parameters for a sample of 57 galaxies, for which 
19 had an accurate $M_{\rm BH}$ measurement and the remaining 38 had only an upper limit 
which are not used here.
They performed 2D decompositions with GASP2D \citep{men2008}, using a S\'ersic 
profile to model the elliptical galaxies and a combination of a S\'ersic plus an exponential model for the disc
galaxies. Galaxies affected by poor decomposition due to either a central bar, a Freeman Type II disc profile \citep{fre1970},
or just inadequately represented by the single or double component modelling were eliminated from their initial sample.
Among their correlations involving the SMBH mass and the parameters of the
host galaxy, the tightest was with the stellar velocity dispersion. 
Little or no correlation was found with the S\'ersic index (see their Figure 7). \\

Table \ref{tab:data} reports the S\'ersic index measurements from the above four works,
along with the type of photometric decomposition performed.
It comprises 62 galaxies. Each galaxy can have up to four S\'ersic index estimates.
35 galaxies have multiple measurements of their S\'ersic index.
In the next two Sections we discuss how we compare and combine them.

  \begin{table*}
  \caption[]{SMBH galaxy sample. 
  Column (1): Galaxy names; 8 galaxies marked with a $*$ have been excluded from the final analysis due to the large disagreement 
  on their S\'ersic index measurements, according to the criteria mentioned in Section \ref{sec:aver}. 
  Column (2): morphological type as listed by \citet{gra2013}, primarily from NED. 
  Column (3): Absolute total B-band magnitudes, from the RC3 catalog 
  using the galaxy distances published in \citet{gra2013}. 
  Column (4): Black hole masses from \citet{gra2013}. 
  Column (5): Presence of a partially depleted core as listed by \citet{gra2013} and such that 
  the question mark is used when the classification has come from the velocity dispersion
  criteria mentioned in Section \ref{sec:res}. 
  Column (6-9): Galaxy decomposition performed by the four works described in Section \ref{sec:ndata}; 
  B = S\'ersic profile, D = disc, g = Gaussian, m = central mask,
  b = bar, p = PSF. 
  Column (10-13): measured S\'ersic index values. }
  \begin{center}
  \begin{tabular}{llrrcccccccrrrr}
 \hline
  Galaxy & Type & $M_{\rm B_{\rm T}}$ & $M_{\rm BH}$ & core & & \multicolumn{4}{c}{Decomposition} & &  \multicolumn{4}{c}{$n$} \\
   &   &  &  &  & & GD07$^{a}$ & V12$^{b}$ &  S11$^{c}$ & B12$^{d}$ &  & GD07$^{a}$ & V12$^{b}$ &  S11$^{c}$ & B12$^{d}$ \\
   &  &  [mag] & $[10^8 \rm M_\odot]$ &  &  &  &  &  &  &  &  &  &  &  \\
  (1) & (2) & (3) & (4) & (5) & & (6) & (7) & (8) & (9) & & (10) & (11) & (12) & (13)\\
  \hline
  Abell 1836-BCG   &  E1     &  $-21.43$  &  $    39	^{+    4  }_{-     5  }$  & y?  &  &	  &	     &    	& BD  &  &	 &	&	  & 2.73 \\			  
  Circinus    	   &  Sb     &  $-15.14$  &  $  0.011	^{+ 0.002 }_{-  0.002 }$  & n?  &  &	  &	     & BD	&      &  &	 &	&   2.0   &	 \\			  
      IC 1459 	   &  E      &  $-21.30$  &  $    24	^{+   10  }_{-    10  }$  & y	&  &	  &	     &  Bg      &      &  &	 &	&   6.0   &	 \\	     
      IC 2560 	   &  SBb    &  $-20.52$  &  $  0.044	^{+ 0.044 }_{-  0.022 }$  & n?  &  &	  &	     & BDg      &      &  &	 &	&   2.0   &	 \\	      
  MESSIER 32  	   &  S0?    &  $-15.46$  &  $  0.024	^{+ 0.005 }_{-  0.005 }$  & n	&  & BD   & BDm      & BD	&      &  & 1.51 & 2.1  &   4.0   &	 \\			  
  MESSIER 59  	   &  E      &  $-20.68$  &  $    3.9	^{+   0.4 }_{-    0.4 }$  & n	&  &	  & Bm	     &  B 	&      &  &	 & 5.7  &   5.0   &	 \\			  
  MESSIER 60  	   &  E1     &  $-21.26$  &  $     47	^{+    10 }_{-     10 }$  & y	&  &  B   & Bm	     & BD	& BD   &  & 6.04 & 3.6  &   3.0   & 1.63 \\			  
  MESSIER 64  	   &  Sab    &  $-19.96$  &  $  0.016	^{+ 0.004 }_{-  0.004 }$  & n?  &  &	  &	     &    	& BD   &  &	 &	&	  & 1.49 \\			  
  MESSIER 77  	   &  SBb    &  $-21.30$  &  $ 0.084	^{+ 0.003 }_{-  0.003 }$  & n	&  &	  & BDbm     & BDg      & BD   &  &	 & 0.8  &   1.0   & 1.27 \\	      
  MESSIER 81  	   &  Sab    &  $-20.01$  &  $   0.74	^{+  0.21 }_{-   0.11 }$  & n	&  & BD   &	     & BDg      & BD   &  & 3.26 &	&   3.0   & 2.57 \\	      
  MESSIER 84  	   &  E1     &  $-21.17$  &  $      9	^{+   0.9 }_{-    0.8 }$  & y	&  &  B   & Bm	     &  Bg      &  B   &  & 5.60 & 3.5  &   7.0   & 4.10 \\	     
  MESSIER 87 *	   &  E0     &  $-21.38$  &  $     58	^{+   3.5 }_{-    3.5 }$  & y?  &  &  B   & Bm	     &  Bg      &      &  & 6.86 & 2.4  &   4.0   &	 \\	     
  MESSIER 89  	   &  E      &  $-20.14$  &  $    4.7	^{+   0.5 }_{-    0.5 }$  & y	&  &	  &   B      & BDg      &  B   &  &	 & 3.6  &   4.0   & 4.30 \\	      
  MESSIER 96  	   &  SBab   &  $-19.91$  &  $  0.073	^{+ 0.015 }_{-  0.015 }$  & n	&  &	  &	     & BDb	&      &  &	 &	&   1.0   &	 \\		  
  MESSIER 104 	   &  Sa     &  $-20.91$  &  $    6.4	^{+   0.4 }_{-    0.4 }$  & y	&  &	  &	     & BDbg     &      &  &	 &	&   1.5   &	 \\		
  MESSIER 105 	   &  E1     &  $-19.82$  &  $      4	^{+	1 }_{-      1 }$  & y	&  &  B   &	     &  B 	&      &  & 4.29 &	&   5.0   &	 \\			  
  MESSIER 106 	   &  SBbc   &  $-20.19$  &  $   0.39	^{+  0.01 }_{-   0.01 }$  & n	&  & BD   & BDp      & BDg      &      &  & 2.04 & 3.5  &   2.0   &	 \\	      
    Milky Way 	   &  SBbc   &  	  &  $   0.043  ^{+ 0.004 }_{-  0.004 }$  & n	&  & BD   &	     &    	&      &  & 1.32 &	&	  &	 \\			  
    NGC 0524  	   &  S0     &  $-20.54$  &  $    8.3	^{+   2.7 }_{-    1.3 }$  & y	&  &	  &	     & BD	&      &  &	 &	&   3.0   &	 \\			  
    NGC 0821  	   &  E      &  $-20.18$  &  $   0.39	^{+  0.26 }_{-   0.09 }$  & n	&  &  B   &	     &  B 	&  B   &  & 4.00 &	&   7.0   & 7.70 \\			  
    NGC 1023  	   &  SB0    &  $-19.88$  &  $   0.42	^{+  0.04 }_{-   0.04 }$  & n	&  & BD   &	     & BDb	&      &  & 2.01 &	&   3.0   &	 \\		  
    NGC 1300  	   &  SBbc   &  $-20.47$  &  $   0.73	^{+  0.69 }_{-   0.35 }$  & n	&  &	  &	     & BD	&      &  &	 &	&   3.0   &	 \\			  
    NGC 1316  	   &  SB0    &  $-21.93$  &  $    1.5	^{+  0.75 }_{-    0.8 }$  & y?  &  &	  &	     & BDg      &      &  &	 &	&   5.0   &	 \\	      
    NGC 1399  	   &  E      &  $-20.89$  &  $    4.7	^{+   0.6 }_{-    0.6 }$  & y	&  &  B   &	     &    	&      &  & 16.8 &	&	  &	 \\			  
    NGC 2549  	   &  SB0    &  $-18.26$  &  $   0.14	^{+  0.02 }_{-   0.13 }$  & n	&  &	  &	     & BD	&      &  &	 &	&   7.0   &	 \\			  
    NGC 2778  	   &  SB0    &  $-18.39$  &  $   0.15	^{+  0.09 }_{-    0.1 }$  & n	&  & BD   &  BD      & BD	&      &  & 1.60 & 2.7  &   2.5   &	 \\			  
    NGC 2787 *	   &  SB0    &  $-17.50$  &  $    0.4	^{+  0.04 }_{-   0.05 }$  & n	&  & BD   &	     & BDbg     &      &  & 1.97 &	&   3.0   &	 \\		
    NGC 2960  	   &  Sa?    &  $-21.25$  &  $   0.12	^{+ 0.005 }_{-  0.005 }$  & n?  &  &	  &  BD      &    	&      &  &	 & 4.0  &	  &	 \\			  
    NGC 2974  	   &  E      &  $-19.73$  &  $    1.7	^{+   0.2 }_{-    0.2 }$  & n	&  &	  &	     &  Bg      &      &  &	 &	&   3.0   &	 \\	     
    NGC 3079  	   &  SBc    &  $-20.04$  &  $  0.024	^{+ 0.024 }_{-  0.012 }$  & n?  &  &	  &	     & BDbg     &      &  &	 &	&   2.0   &	 \\		
    NGC 3115 *	   &  S0     &  $-20.00$  &  $    8.8	^{+    10 }_{-    2.7 }$  & n	&  & BD   &	     & BD	&      &  & 13.0 &	&   3.0   &	 \\			  
    NGC 3227  	   &  SBa    &  $-20.44$  &  $   0.14	^{+   0.1 }_{-   0.06 }$  & n	&  &	  &	     & BD	&      &  &	 &	&   4.0   &	 \\			  
    NGC 3245  	   &  S0     &  $-19.84$  &  $      2	^{+   0.5 }_{-    0.5 }$  & n	&  & BD   &  BD      & BD	& BD  &  & 4.31 & 2.6  &   2.5   & 1.60 \\			  
    NGC 3377  	   &  E5     &  $-18.95$  &  $   0.77	^{+  0.04 }_{-   0.06 }$  & n	&  &  B   &	     &  B 	&  B   &  & 3.04 &	&   6.0   & 3.47 \\			  
    NGC 3384  	   &  SB0    &  $-19.42$  &  $   0.17	^{+  0.01 }_{-   0.02 }$  & n	&  & BD   &	     & BDb	& BD  &  & 1.72 &	&   2.5   & 2.33 \\		  
    NGC 3414  	   &  S0     &  $-19.99$  &  $    2.4	^{+   0.3 }_{-    0.3 }$  & n	&  &	  &	     & BDb	&      &  &	 &	&   5.0   &	 \\		  
    NGC 3489  	   &  SB0    &  $-19.22$  &  $  0.058	^{+ 0.008 }_{-  0.008 }$  & n	&  &	  &	     & BD	&      &  &	 &	&   1.5   &	 \\			  
    NGC 3585  	   &  S0     &  $-20.57$  &  $    3.1	^{+   1.4 }_{-    0.6 }$  & n	&  &	  &	     & BD	&      &  &	 &	&   2.5   &	 \\			  
    NGC 3607  	   &  S0     &  $-20.91$  &  $    1.3	^{+   0.5 }_{-    0.5 }$  & n	&  &	  &	     &  Bg      &  B   &  &	 &	&   5.0   & 4.70 \\	     
    NGC 3608 *	   &  E2     &  $-20.04$  &  $      2	^{+   1.1 }_{-    0.6 }$  & y	&  &	  &	     &  B 	&  B   &  &	 &	&   6.0   & 9.03 \\			  
    NGC 3998 *	   &  S0     &  $-19.07$  &  $    8.1	^{+	2 }_{-    1.9 }$  & y?  &  &	  &	     & BDg      & BD  &  &	 &	&   1.5   & 2.29 \\	      
    NGC 4026  	   &  S0     &  $-18.93$  &  $    1.8	^{+   0.6 }_{-    0.3 }$  & n	&  &	  &	     & BD	&      &  &	 &	&   3.5   &	 \\			  
    NGC 4151  	   &  SBab   &  $-20.01$  &  $   0.65	^{+  0.07 }_{-   0.07 }$  & n	&  &	  &	     & BDg      &      &  &	 &	&   3.5   &	 \\	      
    NGC 4261  	   &  E2     &  $-21.03$  &  $      5	^{+	1 }_{-      1 }$  & y	&  &  B   & Bm	     & BDg      &  B   &  & 7.30 & 3.5  &   4.0   & 4.31 \\	      
    NGC 4291  	   &  E2     &  $-19.60$  &  $    3.3	^{+   0.9 }_{-    2.5 }$  & y	&  &  B   &	     &    	&      &  & 4.02 &	&	  &	 \\			  
    NGC 4342 *	   &  S0     &  $-18.40$  &  $    4.5	^{+   2.3 }_{-    1.5 }$  & n	&  & BD   &  BD      &    	&      &  & 5.11 & 1.9  &	  &	 \\			  
    NGC 4459 *	   &  S0     &  $-19.66$  &  $   0.68	^{+  0.13 }_{-   0.13 }$  & n	&  &	  & B        & BD	&  B   &  &	 & 3.9  &   2.5   & 7.44 \\			  
  \hline
  \end{tabular}
  \end{center}
  \label{tab:data}
  \end{table*}

  \begin{table*}
  \begin{center}
  \begin{tabular}{llrrcccccccrrrr}
  \multicolumn{12}{l}{{\bf Table \ref{tab:data}.} \emph{Continued.}} \\
  \hline
  Galaxy & Type & $M_{\rm B_{\rm T}}$ & $M_{\rm BH}$ & core & & \multicolumn{4}{c}{Decomposition} & &  \multicolumn{4}{c}{$n$} \\
   &   &  &  &  & & GD07$^{a}$ & V12$^{b}$ &  S11$^{c}$ & B12$^{d}$ &  & GD07$^{a}$ & V12$^{b}$ &  S11$^{c}$ & B12$^{d}$ \\
   &  &  [mag] & $[10^8 \rm M_\odot]$ &  &  &  &  &  &  &  &  &  &  &  \\
  (1) & (2) & (3) & (4) & (5) & & (6) & (7) & (8) & (9) & & (10) & (11) & (12) & (13)\\
  \hline
    NGC 4473  	   &  E5     &  $-19.76$  &  $    1.2	^{+   0.4 }_{-    0.9 }$  & n	&  &  B   & Bm	     &  B 	& BD   &  & 2.73 & 4.3  &   7.0   & 2.23 \\			  
    NGC 4486A 	   &  E2     &  $-18.04$  &  $   0.13	^{+  0.08 }_{-   0.08 }$  & n	&  &	  & Bm	     &  B 	&      &  &	 & 2.0  &   2.5   &	 \\			  
    NGC 4564  	   &  S0     &  $-18.77$  &  $    0.6	^{+  0.03 }_{-   0.09 }$  & n	&  & BD   &  BD      & BD	&      &  & 3.15 & 3.7  &   7.0   &	 \\			  
    NGC 4596  	   &  SB0    &  $-19.80$  &  $   0.79	^{+  0.38 }_{-   0.33 }$  & n	&  &	  & BDb      & BDb	& BD   &  &	 & 3.6  &   3.0   & 4.43 \\		  
    NGC 4697  	   &  E4     &  $-20.14$  &  $    1.8	^{+   0.2 }_{-    0.1 }$  & n	&  &  B   &   B      &  B 	&  B   &  & 4.00 & 3.8  &   5.0   & 4.96 \\			  
    NGC 5077  	   &  E3     &  $-20.69$  &  $    7.4	^{+   4.7 }_{-      3 }$  & y	&  &	  &	     &  Bg      &      &  &	 &	&   6.0   &	 \\	     
    NGC 5128  	   &  S0     &  $-20.06$  &  $   0.45	^{+  0.17 }_{-    0.1 }$  & n?  &  &	  &	     & BDbg     &      &  &	 &	&   3.5   &	 \\		
    NGC 5252  	   &  S0     &  $-21.03$  &  $     11	^{+    16 }_{-      5 }$  & n	&  &	  &	     &    	& BD   &  &	 &	&	  & 4.82 \\			  
    NGC 5576  	   &  E3     &  $-20.12$  &  $    1.6	^{+   0.3 }_{-    0.4 }$  & n	&  &	  & Bm	     &  B 	&  B   &  &	 & 5.1  &   7.0   & 8.71 \\			  
    NGC 5813  	   &  E      &  $-21.03$  &  $    6.8	^{+   0.7 }_{-    0.7 }$  & y	&  &	  & Bm	     &  B 	&      &  &	 & 8.3  &   6.0   &	 \\			  
    NGC 5845  	   &  E3     &  $-18.51$  &  $    2.6	^{+   0.4 }_{-    1.5 }$  & n	&  &  B   &   B      &  B 	&  B   &  & 3.22 & 2.6  &   3.0   & 3.45 \\			  
    NGC 5846  	   &  E      &  $-20.87$  &  $     11	^{+	1 }_{-      1 }$  & y	&  &	  & Bm	     &  B 	&      &  &	 & 3.7  &   3.0   &	 \\			  
    NGC 6251 *	   &  E2     &  $-21.46$  &  $    5.9	^{+	2 }_{-      2 }$  & y?  &  &  B   &	     &  Bg      &      &  & 11.8 &	&   7.0   &	 \\	     
    NGC 7052  	   &  E      &  $-20.71$  &  $    3.7	^{+   2.6 }_{-    1.5 }$  & y	&  &  B   &  BD      &  B 	&      &  & 4.55 & 1.8  &   5.0   &	 \\			  
    NGC 7582  	   &  SBab   &  $-20.34$  &  $   0.55	^{+  0.26 }_{-   0.19 }$  & n	&  &	  &	     & BDg      &      &  &	 &	&   4.0   &	 \\	      
  \hline
  \multicolumn{12}{l}{$^{a}$ \citet{gra2007}. $^{b}$ \citet{vik2012}. $^{c}$ \citet{san2011}. $^{d}$ \citet{bei2012}.} \\
  \end{tabular}
  \end{center}
  \end{table*}

\subsection{Comparing S\'ersic indices}
\label{sec:disag}
There are three main points that distinguish each study:  
the first is the wavelength of the image. 

The spatial distribution of the surface brightness of a galaxy, and hence
its light profile, is a function of the observational bandpass. This means that the structural parameters, in general,
may vary with wavelength due to stellar population gradients or dust
obscuration. The central light concentration of a galaxy, described by the S\'ersic index, is indeed
a slight function of wavelength. 
Using reprocessed Sloan Digital Sky Survey Data Release Seven (SDSS DR7, \citealt{dr7}) and UKIRT Infrared Deep Sky Survey
Large Area Survey \citep{ukirt} imaging data available from the Galaxy And Mass Assembly (GAMA) database, 
\citet{kel2012} performed 2D model fits with GALFIT to $\sim$170,000 galaxies in the $ugrizYJHK$ bandpasses, using primarily
a pure S\'ersic profile, to quantify how photometric and structural parameters of a galaxy vary with wavelength.
Their Figure 21 shows the mean S\'ersic index as a function of the rest-frame wavelength for two subsamples: the disc-dominated
and the spheroid-dominated systems. \citet{kel2012} find that the spheroid-dominated population is characterized by
mean S\'ersic indices that remain relatively stable at all wavelengths, with $n$ increasing by 30\% from $g$ to $K$.

The second point is the model fitting method: one-dimensional and two-dimensional photometric decomposition techniques,
if performed on the same galaxy, can produce different values of the S\'ersic index due to ellipticity gradients
which the 2D models cannot accommodate. 
The parameters of the S\'ersic model can vary if derived along the major- or 
the minor-axis, as first noted by \citet{cao1993}.
\citet{fer2004} quantified such discrepancy in terms of ellipticity gradients, 
i.e. the isophote eccentricity that varies with radius.
The histogram in Figure \ref{fig:caon} has been created using data 
from \citet{cao1993} and shows the distribution of the ratio between the 
``equivalent'' S\'ersic index $n_{\rm eq}$ and that measured along the major-axis, $n_{\rm maj}$.
The ``equivalent'' axis is the geometric mean, $\sqrt{ab}$, of the major and the minor 
axis of the isophotal ellipses.
The mean (and the standard deviation) of the whole sample is 
$<n_{\rm eq}/n_{\rm maj}> = 1.10 \pm 0.27$.
This tells us that the equivalent S\'ersic index is on average 10\% higher 
than the major-axis S\'ersic index.
From Figure \ref{fig:caon}, their relative
difference will be less than 40\% in 95\% of the time. 

The third issue pertains to the weighting-scheme used for the fits.
The arrival of photons, which build up a galaxy image, is a Poissonian process ($noise \propto \sqrt{signal}$),
which therefore advocates the need for a signal-to-noise weighted fitting scheme.
However the presence of active galactic nuclei, nuclear star clusters, nuclear stellar discs, 
dust, partially-depleted cores and an uncertain PSF make such a weighting 
prone to error unless all of these factors are taken into account.

Hence, what do we expect from our heterogeneous collection of data? 
First, the wavelength bias should produce a systematic effect in the S\'ersic index measurements, 
i.e. we expect the measurements from GD07 (R-band) and 
B12 ($i$-band) to be slightly smaller than those from V12 (K-band) and S11 ($3.6~\mu$m). 
Second, because the S\'ersic index derived from a two-dimensional analysis can be approximated
to the one-dimensional $n_{\rm eq}$, one may expect 
the S\'ersic index derived from one-dimensional decomposition along the major-axis,
as performed by GD07, to be slightly smaller than the S\'ersic 
index derived from the two-dimensional modelling in V12, S11 and B12.
However, when we compare different measurements of the S\'ersic index 
(belonging to the same galaxy), we do not observe the previous
systematic effects; moreover, for a non negligible number of galaxies 
we find that multiple measurements have a relative 
difference\footnote{Given two measurements $n_1$ and $n_2$, with $n_1 < n_2$,
we define the \emph{relative difference} as $(n_2 - n_1)/n_1$.} greater than 50\%.

\begin{table*}
  \centering
  \begin{minipage}{120mm}
    \caption{Examples of outlying measurements, used to explain the crossed out data in Figure \ref{fig:bmagnzoom}. 
    Column (1): Galaxy names. 
    Columns (2,4,6,8): Literature S\'ersic index measurements in ascending order; 
    the reference is given in the superscript. 
    Columns (3,5,7): Relative differences; bold type is used for values greater than 50\%.}
    \label{tab:out}
    \begin{tabular}{llclclcl}
       \hline
       Galaxy	& $n_1$ 	     & $\frac{n_2-n_1}{n_1}$   & $n_2$  	   & $\frac{n_3-n_2}{n_2}$    & $n_3$		    & $\frac{n_4-n_3}{n_3}$    & $n_4$  	   \\
       (1) & (2) & (3) & (4) & (5) & (6) & (7) & (8) \\
       \hline
       \multicolumn{8}{c}{\emph{Galaxies with same choice of decomposition}} \\
       M87	& $2.4^{\rm V12}$    & $\bf 0.67$	       & $4.0^{\rm S11}$   & $\bf 0.72$ 	      & $6.86^{\rm GD07}$   &			       &		   \\
       NGC 0821 & $4.0^{\rm GD07}$   & $\bf 0.75$              & $7.0^{\rm S11}$   & $0.10$                   & $7.70^{\rm B12}$    &			       &		   \\
       NGC 3115 & $3.0^{\rm S11}$    & $\bf 3.33$	       & $13.0^{\rm GD07}$ &			      & 		    &			       &		   \\
       NGC 4342 & $1.9^{\rm V12}$    & $\bf 1.69$	       & $5.11^{\rm GD07}$ &			      & 		    &			       &		   \\
       NGC 4564 & $3.15^{\rm GD07}$  & $0.17$		       & $3.7^{\rm V12}$   & $\bf 0.89$ 	      & $7.0^{\rm S11}$     &			       &		   \\
       NGC 6251 & $7.0^{\rm S11}$    & $\bf 0.69$	       & $11.8^{\rm GD07}$ &			      & 		    &			       &		   \\
       \hline
       \multicolumn{8}{c}{\emph{Galaxies with different choices of decomposition}} \\
       M60	& $1.63^{\rm B12}$   & $\bf 0.84$	       & $3.0^{\rm S11}$   & $0.20$		      & $3.6^{\rm V12}$     & $\bf 0.68$	       & $6.04^{\rm GD07}$ \\
       NGC 4459 & $2.5^{\rm S11}$    & $\bf 0.56$	       & $3.9^{\rm V12}$   & $\bf 0.91$ 	      & $7.44^{\rm B12}$    &			       &		   \\
       NGC 7052 & $1.8^{\rm V12}$    & $\bf 1.53$	       & $4.55^{\rm GD07}$ & $0.10$		      & $5.0^{\rm S11}$     &			       &		   \\
       \hline
  \end{tabular}
  \end{minipage}
\end{table*}

Many factors, if not properly taken into account, can affect the model-fitting of the light distribution 
of a galaxy and hence the derivation of its structural parameters.
These factors can include:
additional nuclear components; 
the presence of a bar; a partially depleted core in high resolution images; a bad sky subtraction, etc. 
Moreover, different choices of structural components for the same galaxy will produce 
contrasting S\'ersic indices.
Table \ref{tab:out} reports a few examples of discrepant measurements.
For the first five galaxies, each study used the same type of decomposition (S\'ersic 
or S\'ersic+exponential). For the last three galaxies each study performed a different 
image decomposition. M60 was modelled with a pure S\'ersic profile by GD07 and V12,
while S11 and B12 used an additional disc component. 
NGC 4459 has a bulge+disc profile according to S11, while V12 and B12 agreed in
modelling the galaxy with a pure S\'ersic profile.
GD07 and S11 fit NGC 7052 with a pure S\'ersic profile, whereas V12 chose a bulge+disc model.
An exhaustive analysis of why the individual S\'ersic indices differ from author to author 
is however beyond the scope of the present work.

\subsection{Combining S\'ersic indices}
\label{sec:aver}
To combine the results of these four heterogeneous works,
we decided to use a method
that was as simple as possible and that involved the least manipulation of the data. 
Our strategy consisted of looking at galaxies with multiple measurements, comparing the different S\'ersic indices
and excluding the most contrasting measurements before then 
averaging the remaining S\'ersic indices. 

The exclusion-algorithm is the following: given a galaxy A that has been analyzed by more than one study,
we take each measurement $n_{\rm i}^{\rm A}$ and we look for the closest one $n_{\rm j}^{\rm A}$.
If the absolute difference $|\Delta n_{\rm ij}^{\rm A}| = |n_{\rm i}^{\rm A} - n_{\rm j}^{\rm A}|$ 
is more than 50\% of the minimum among the two measurements, 
we exclude $n_{\rm i}^{\rm A}$. Obviously, if a galaxy has only two measurements, 
we exclude both of them.  
After applying the exclusion-algorithm, we compute the average 
logarithmic value of the remaining measurements to give us $\langle\log(n^{\rm A})\rangle$.

\begin{figure}
\begin{center}
\includegraphics[width=\columnwidth, trim=0mm 10mm 10mm 10mm]{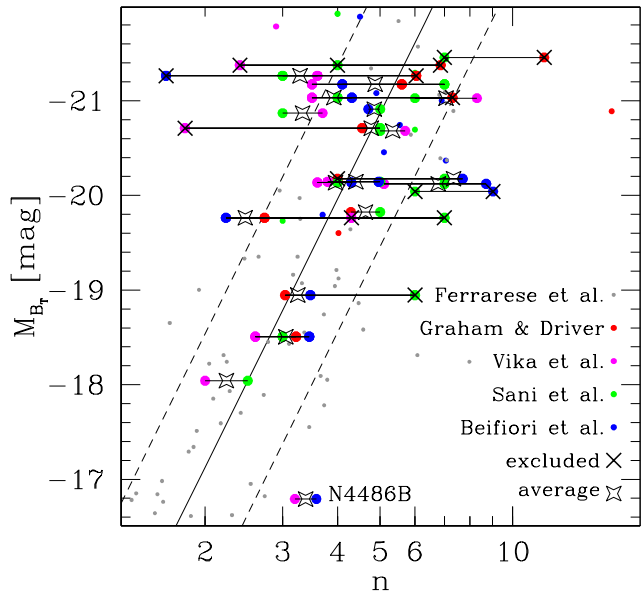}
\end{center}
\caption{Absolute B-band magnitude vs S\'ersic index
of elliptical galaxies. This Figure is a ``zoom'' of the dotted box in Figure \ref{fig:bmagn} and
it uses the same colour coding (see the previous caption). 
The black solid line and the dashed lines are again the $M_{\rm B_{\rm T}}-n$ relation from \citet{gra2003a}
and a rough ``by eye'' estimate of the scatter in their diagram. 
The grey points are excluded from the following description.
Horizontal solid lines connect different S\'ersic measurements of the same galaxy.
Bigger dots refer to galaxies with multiple measurements, while smaller dots show
galaxies with only one measurement. 
Black crosses mark the excluded measurements, according to the algorithm 
described in Section \ref{sec:aver} and illustrated in Table \ref{tab:out}.
Big empty stars indicate the average S\'ersic index $\langle \log(n) \rangle$ derived from the 
``good'' (not excluded) S\'ersic measurements.}
\label{fig:bmagnzoom}
\end{figure}

Figures \ref{fig:bmagn} and \ref{fig:bmagnzoom} are helpful to visualize
our approach. 
Figure \ref{fig:bmagnzoom} is a ``zoom'' of Figure \ref{fig:bmagn}
and they both show the absolute total B-band magnitude $M_{\rm B_{\rm T}}$
of elliptical galaxies plotted against their S\'ersic index.
The black solid line shows the $M_{\rm B_{\rm T}}-n$ relation from \citet{gra2003a} 
such that $M_{\rm B_{\rm T}} = -9.4 \log(n) - 14.3$,
while the dashed lines are a rough ``by eye'' estimate of its scatter.

The horizontal solid lines in Figure \ref{fig:bmagnzoom} connect the different S\'ersic index measurements
of the same galaxy.  If a galaxy's S\'ersic index has been measured by more than one
study, it is represented with a bigger dot. Thus, small dots
refer to galaxies that have been measured by only one study.
A black cross on a dot means that we intend to exclude that
particular measurement because it is in strong disagreement ($>50\%$) with the
other points according to our exclusion-algorithm. The average $\langle\log(n^{\rm A}) \rangle$
of the logarithmic values of the remaining measurements 
is denoted by an empty star. 

We apply the same procedure to the bulges of the lenticular and spiral
galaxies, which are not shown in the $M_{\rm B_{\rm T}}-n$ plots (Figures \ref{fig:bmagn} and \ref{fig:bmagnzoom}), 
but are included in the following analysis. 
Our final sample consists of 54 galaxies with directly measured SMBH mass and 
at least one measurement of the S\'ersic index; among these, 27 galaxies have indices measured by more than one 
study. The 8 galaxies excluded from the initial sample of 62 objects,
due to widely varying S\'ersic indices, are marked with a star
in Table \ref{tab:data}.

\section{Analysis}
\label{sec:res}
\begin{figure*}
\begin{center}
\includegraphics[width=2.2\columnwidth, trim= 10mm 20mm 0mm 90mm]{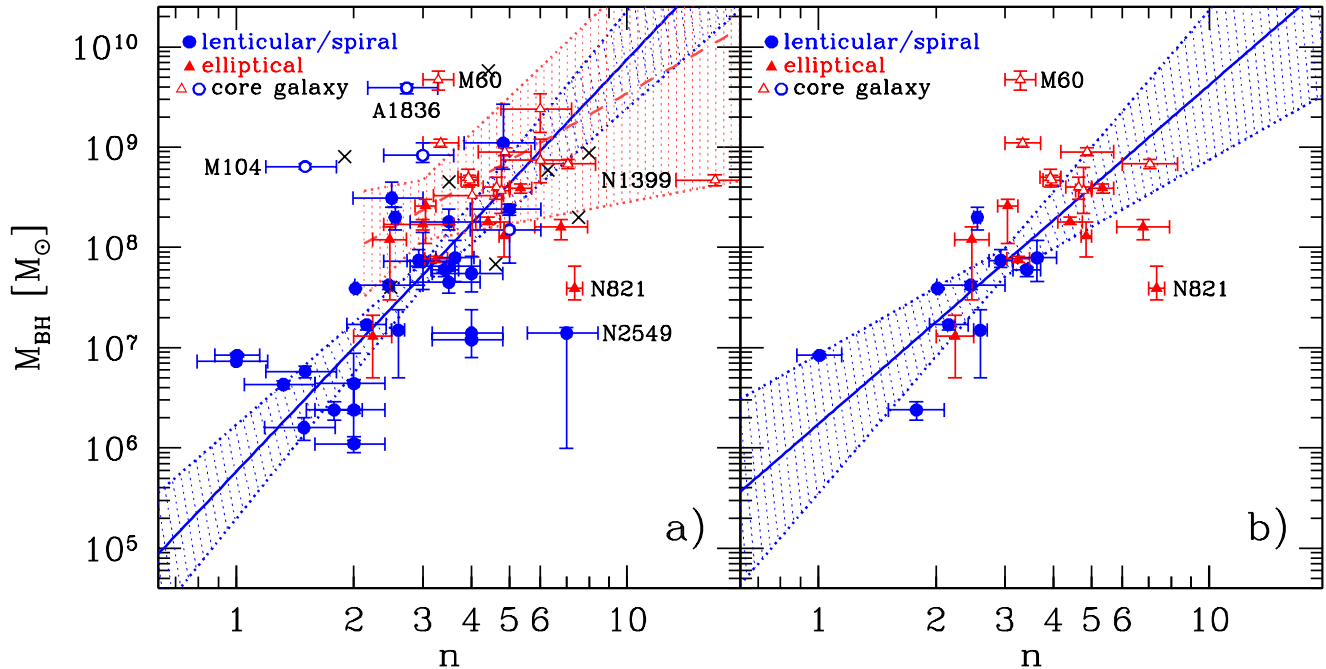}
\end{center}
\caption{Black hole mass vs S\'ersic index.
{\bf Figure \ref{fig:mbhn_fit}a}: All galaxies with at least one measurement from GD07,
V12, S11 and B12; 
if a galaxy has been measured by more than one study, we plot the average
value of its S\'ersic index as obtained in Section \ref{sec:aver}. 
Black crosses are used to show the location of the 8 galaxies excluded from the initial sample of 62,
due to widely varying S\'ersic indices (we plot their mean S\'ersic index).
{\bf Figure \ref{fig:mbhn_fit}b}: Only the 27 galaxies with multiple S\'ersic measurements. 
Open symbols are used for core-S\'ersic galaxies,
rather than filled symbols that denote S\'ersic galaxies.
The solid blue line (and the blue dotted lines) shows the symmetrical bisector regression (with errors) for
the S\'ersic bulges of disc galaxies.
The dashed red line (and the red dotted lines) shows the symmetrical bisector regression (with errors) for
core-S\'ersic elliptical galaxies (not shown in Figure \ref{fig:mbhn_fit}b due to the low number statistics).
The labelled galaxies designate outliers that were excluded from the regressions.}
\label{fig:mbhn_fit}
\end{figure*}

After taking galaxies with multiple S\'ersic index measurements, 
rejecting the outlying values and averaging
the remaining ones, according to
the strategy discussed in Section \ref{sec:aver},
we build the $M_{\rm BH} - n$ diagram. 
For galaxies with multiple measurements, we calculated
the error on their mean S\'ersic index,
whereas for single-measured objects we assumed an
error\footnote{The error of single-measured objects
was estimated as follows. Using the 35 galaxies with multiple measurements of their
S\'ersic indices, we first computed the average $\langle \log(n) \rangle$ of each galaxy without applying
the exclusion algorithm (see Section \ref{sec:aver}) and its error $\sigma_{\langle \log(n) \rangle}$;
we then calculated the median value of the errors $\langle \sigma_{\langle \log(n) \rangle} \rangle = 0.08$ (20\%).}
of 20\%. 
Figure \ref{fig:mbhn_fit}a includes galaxies with single and averaged-multiple S\'ersic indices,
whereas Figure \ref{fig:mbhn_fit}b only shows those with an averaged-multiple measurement 
and is thus more reliable.

Despite the higher level of scatter in Figure \ref{fig:mbhn_fit}a, 
both diagrams display an appreciable correlation
between the SMBH mass and the spheroid light concentration.
That is, after excluding the discrepant S\'ersic indices
according to the process in Section \ref{sec:aver}, presumably from poor fits,
we recover a clear trend between black hole mass and S\'ersic index.
We have visually identified six\footnote{Abell 1836-BCG, M60, M104, NGC 1399, NGC 821, NGC 2549.} outliers
in Figure \ref{fig:mbhn_fit}a and two\footnote{M60, NGC 821.} outliers in Figure \ref{fig:mbhn_fit}b;
these objects are labelled in both diagrams and were excluded from the following regression analysis.
The Spearman's correlation coefficients $r_{\rm s}$ and the likelihood of the correlation occuring by chance $P$ 
are given in Table \ref{tab:spear}. 
In both panels we have performed a symmetrical linear 
bisector regression  
using the BCES routine from \citet{akr1996},
which was checked using the Bayesian linear regression code \emph{linmix\_err} \citep{kel2007}. 
However, we have not lumped all the galaxy data together, as there is good reason 
not to do this. 

\begin{table}
  \caption{Spearman's correlation coefficients $r_{\rm s}(N-2)$ and likelihood of the correlation occuring by chance $P$.
  $N-2$ are the degrees of freedom.}
  \label{tab:spear}
   \begin{center}
    \begin{tabular}{lll}
       \hline
       Figure \ref{fig:mbhn_fit}a excluding outliers & $r_{\rm s}(46) = 0.72$ & $P < 0.1\%$ \\
       Figure \ref{fig:mbhn_fit}a including outliers & $r_{\rm s}(52) = 0.53$ & $P < 0.1\%$ \\
       Figure \ref{fig:mbhn_fit}b excluding outliers & $r_{\rm s}(23) = 0.76$ & $P < 0.1\%$ \\
       Figure \ref{fig:mbhn_fit}b including outliers & $r_{\rm s}(25) = 0.60$ & $P < 1\%$ \\
       \hline
    \end{tabular}
   \end{center}
\end{table}

Among our galaxy sample with direct $M_{\rm BH}$ measurements, 
\citet{gra2013} identified ``core-S\'ersic'' galaxies that display 
a central deficit of light relative to the inward extrapolation of their outer S\'ersic 
light profile, and ``S\'ersic'' galaxies that do not \citep{gra2003a,gra2003b,tru2004}.
``Core-S\'ersic'' galaxies are thought to have formed from dry merger events, 
whereas ``S\'ersic'' galaxies are the result of gaseous processes.
Their classification (Column 5 of Table \ref{tab:data})
has primarily come from the inspection of high-resolution images. When no core designation
was available or possible from the literature, \citet{gra2013} used a criteria based on the velocity
dispersion $\sigma$, such that galaxies with $\sigma > 270 \rm ~km~s^{-1}$ are considered likely to
possess a partially depleted core, while galaxies with $\sigma < 165 \rm ~km~s^{-1}$ are not.
For reasons discussed in Section \ref{sec:predmbhn}, we divided our sample into four subsamples:
\begin{enumerate} 
\item[-] the S\'ersic bulges of disc galaxies;
\item[-] S\'ersic elliptical galaxies;
\item[-] the core-S\'ersic bulges of disc galaxies;
\item[-] core-S\'ersic elliptical galaxies.
\end{enumerate} 
We expect a different $M_{\rm BH} - n$ relation for each of the previous subsamples,
and hence we elect not to perform a single linear regression to all the data 
shown in Figures \ref{fig:mbhn_fit}a and \ref{fig:mbhn_fit}b.
Our symmetrical regressions have been performed for
the S\'ersic bulges of disc galaxies in Figures \ref{fig:mbhn_fit}a and \ref{fig:mbhn_fit}b
and for core-S\'ersic elliptical galaxies in Figure \ref{fig:mbhn_fit}a.
Due to small numbers, the statistics were not able to provide reliable regressions for 
core-S\'ersic elliptical galaxies in Figure \ref{fig:mbhn_fit}b, nor for
S\'ersic elliptical galaxies and core-S\'ersic bulges in either Figures \ref{fig:mbhn_fit}a and
\ref{fig:mbhn_fit}b.
Table \ref{tab:fit} contains the results from the symmetrical regressions.
All of the outliers reside more than $3\sigma$ from the 
linear regressions. \\

\begin{table*}
  \begin{center}
  \begin{minipage}{140mm}
  \caption{Observed $M_{\rm BH} - n$ scaling relations. $M_{\rm BH} =$ black hole mass, 
  $n =$ S\'ersic index.
  A symmetrical bisector regression (BCES routine from \citealt{akr1996}) was used.
  The quantity $n$ is normalized to the round median value of the distribution
  of the S\'ersic indices for the SMBH galaxy sample ($\langle n \rangle = 3$).
  The total rms scatter in the $\log(M_{\rm BH})$ 
  direction is denoted by $\Delta$.}
  \label{tab:fit}
   \begin{center}
  \begin{tabular}{rlccc}
   \hline
   \# & Type  & $\alpha$ & $\beta$ & $\Delta~\rm dex$ \\
    \hline
    \multicolumn{4}{c}{Figure \ref{fig:mbhn_fit}a} \\				   
    \multicolumn{4}{c}{$\log(M_{\rm BH}/{\rm M_\odot}) = \alpha + \beta \log(n/3)$ } \\ 
       &                                   & 		      & 		 &	 \\   
     9 & S\'ersic elliptical galaxies      &  ...	      & ...		 & ...  \\
    27 & S\'ersic bulges                   &  $7.73 \pm 0.12$ & $4.11 \pm 0.72$  & 0.62 \\
    10 & Core-S\'ersic elliptical galaxies &  $8.37 \pm 0.30$ & $2.23 \pm 1.50$  & 0.27 \\
     2 & Core-S\'ersic bulges              &  ...	      & ...		 & ...  \\
       &                                   & 		      & 		 &	 \\   
    \hline									   
    \multicolumn{4}{c}{Figure \ref{fig:mbhn_fit}b} \\				   
    \multicolumn{4}{c}{$\log(M_{\rm BH}/{\rm M_\odot}) = \alpha + \beta \log(n/3)$ } \\ 
       &                                   & 		      &			 &	\\    
     8 & S\'ersic elliptical galaxies	   &  ...	      & ...		 & ...  \\
    10 & S\'ersic bulges        	   &  $7.85 \pm 0.14$ & $3.38 \pm 1.16$  & 0.44 \\
     7 & Core-S\'ersic elliptical galaxies &  ...	      & ...		 & ...  \\
     0 & Core-S\'ersic bulges              &  ...	      & ...		 & ...  \\
       &                                   & 		      &			 &	\\    
   \hline
  \end{tabular}
   \end{center}
  \end{minipage}  
\end{center}
\end{table*}


\section{Predictions and discussion} 
\label{sec:predmbhn}
The $M_{\rm BH} - n$ relation can be predicted from two other important scaling relations:
the $M_{\rm BH} - L_{\rm sph}$ and the $L_{\rm sph} - n$ relations, where $L_{\rm sph}$ is the
luminosity of the galaxy's spheroidal component.

Since at least \citeauthor{gra2001a} (\citeyear{gra2001a}, his Figure 14), we have known that
the $L_{\rm sph} - n$ relation is different for elliptical galaxies and the bulges of disc galaxies.
Figure 10 from \citet{gra2003a} and Figure 11 from \citet{gra2013rev} display the $L_{\rm sph} - n$ 
relation for elliptical galaxies (in the B-band) and for the bulges of disc galaxies
(in the K$_{\rm s}$-band) respectively. 
In both Figures, the linear regressions had been estimated ``by eye''. 
We re-analyzed the data from their Figures and performed a symmetrical linear 
bisector regression analysis using the BCES routine from \citet{akr1996}. \\
We obtained
\begin{equation*}
   M_{\rm B,sph} = (-18.25 \pm 0.18) + (-9.01 \pm 0.47) \log(n/3)
\label{eq:bmagn}
\end{equation*}
for the elliptical galaxies, and
\begin{equation*}
   M_{\rm K_{\rm s},sph} = (-23.01 \pm 0.15) + (-5.55 \pm 0.47) \log(n/3)
\label{eq:kmagn}   
\end{equation*}
for the bulges of the disc galaxies.
Here $M_{\rm B,sph}$ indicates the absolute B-band magnitude of elliptical galaxies
and $M_{\rm K_s,sph}$ indicates the dust-corrected, absolute K$_{\rm s}$-band magnitude of the bulges of disc galaxies.

We have used the $M_{\rm BH} - L_{\rm sph}$ relation from \citet{gra2013}
who derived B-band and K$_{\rm s}$-band bulge magnitudes, from the total luminosity of lenticular and spiral galaxies,
through a statistical correction that takes into account inclination effects and dust absorption.
Following \citet{gra2012}, \citet{gra2013} derived the $M_{\rm BH} - L_{\rm sph}$ relation separately for core-S\'ersic
and S\'ersic spheroids. 
They observed a near-linear $M_{\rm BH} - L_{\rm sph}$ relation for the core-S\'ersic spheroids,
thought to be built in additive dry merger events, and a notably (2.5 times) steeper $M_{\rm BH} - L_{\rm sph}$ relation
for the S\'ersic spheroids considered to be products of gas-rich processes. 
They reported 
\begin{equation*}
   \log(M_{\rm BH}) = (9.03 \pm 0.09) + (-0.54 \pm 0.12) (M_{\rm B,sph} + 21)
\label{eq:mbhbmagcs}
\end{equation*}
and
\begin{equation*}
   \log(M_{\rm BH}) = (9.05 \pm 0.09) + (-0.44 \pm 0.08) (M_{\rm K_s,sph} + 25)
\label{eq:mbhkmagcs}
\end{equation*}
for their core-S\'ersic subsample, whereas
\begin{equation*}
   \log(M_{\rm BH}) = (7.37 \pm 0.15) + (-0.94 \pm 0.16) (M_{\rm B,sph} + 19)
\label{eq:mbhnbmags}
\end{equation*}
and
\begin{equation*}
    \log(M_{\rm BH}) = (7.39 \pm 0.14) + (-1.09 \pm 0.22) (M_{\rm K_s,sph} + 22.5)
\label{eq:mbhnkmags}    
\end{equation*}
for their S\'ersic galaxies. 

\begin{figure}
\begin{center}
\includegraphics[width=\columnwidth, trim= 0mm 10mm 10mm 10mm]{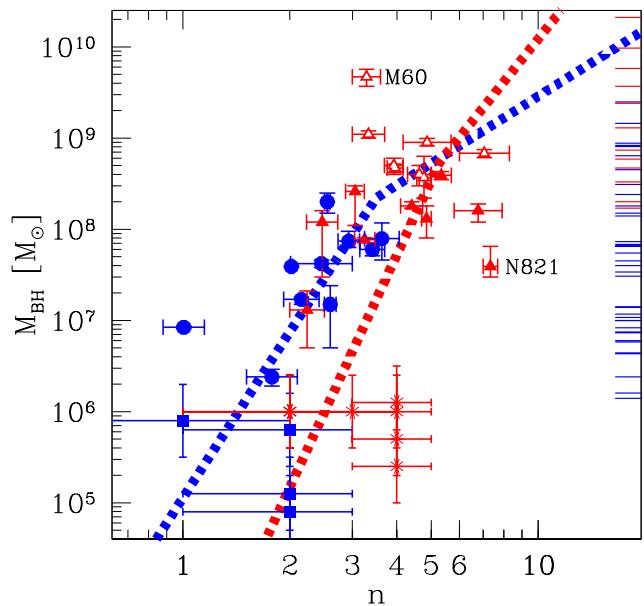}
\end{center}
\caption{Same data as Figure \ref{fig:mbhn_fit}b.
For comparison, we plot 4 additional bulges of disc galaxies (blue squares) 
and 6 additional elliptical galaxies (red asterisks) 
taken from the sample of \citet{gre2008}.
The dashed lines show the \emph{predicted} $M_{\rm BH} - n$ relations for
elliptical galaxies in red and for the bulges of
disc galaxies in blue, given the \emph{observed} $M_{\rm BH} - L_{\rm sph}$ and the $L_{\rm sph} - n$ relations
in the literature.
The ticks on the right axis indicate the black hole masses of 14 elliptical galaxies (in red)
and 37 disc galaxies (in blue) that belong to the sample of \citet{gra2013} 
(and hence have a secure $M_{\rm BH}$ detection) but do not have 
multiple S\'ersic index measurements.}
\label{fig:mbhnpred}
\end{figure}

The \emph{bent} nature of the above $M_{\rm BH} - L_{\rm sph}$ relations 
and the \emph{linear} nature of the two distinct $L_{\rm sph} - n$ relations for elliptical galaxies and bulges
requires that there be two distinct \emph{bent} $M_{\rm BH} - n$ relations for elliptical galaxies and bulges.
This explains the \emph{curved} nature of the $M_{\rm BH} - n$ relation reported by \citet{gra2007}.
The \emph{predicted} $M_{\rm BH} - n$ relations, derived from the above six equations, are reported in Table \ref{tab:pred}
and shown in Figure \ref{fig:mbhnpred}.\\

\begin{table*}
  \begin{center}
  \begin{minipage}{110mm}
  \caption{Predicted $M_{\rm BH} - n$ relations.}
  \label{tab:pred}
   \begin{center}
    \begin{tabular}{ll}
       \hline
       Type & Prediction \\
       \hline
       S\'ersic elliptical galaxies      & $\log(M_{\rm BH}) = (6.66 \pm 0.26) + (8.47 \pm 1.51) \log(n/3)$  \\	      
       S\'ersic bulges                   & $\log(M_{\rm BH}) = (7.95 \pm 0.24) + (6.05 \pm 1.32) \log(n/3)$  \\
       Core-S\'ersic elliptical galaxies & $\log(M_{\rm BH}) = (7.54 \pm 0.35) + (4.87 \pm 1.11) \log(n/3)$  \\  
       Core-S\'ersic bulges              & $\log(M_{\rm BH}) = (8.17 \pm 0.19) + (2.44 \pm 0.49) \log(n/3)$  \\	   
       \hline
    \end{tabular}
   \end{center}
  \end{minipage} 
  \end{center}
\end{table*}

The expected $M_{\rm BH} - n$ relations for the S\'ersic bulges of disc galaxies 
and for core-S\'ersic elliptical galaxies (Table \ref{tab:pred}) are marginally
consistent at the $2\sigma$ level with the results from the linear regression analysis performed in
Figure \ref{fig:mbhn_fit} (Table \ref{tab:fit}).
More quality data and a wider range of S\'ersic indices
would be beneficial to confirm the predicted relations.

For comparison, in Figure \ref{fig:mbhnpred} we plot 10 additional galaxies with $M_{\rm BH} < 10^7 ~\rm M_\odot$
taken from the sample of \citet{gre2008}. 
The horizontal offset that separates the bulges of their 4 disc galaxies from their 6 elliptical galaxies
supports our predicted gap between the $M_{\rm BH} - n$ relations for elliptical galaxies 
and bulges at the low-mass end of this diagram. 
If the \emph{bent} $M_{\rm BH} - L_{\rm sph}$ relation is 
the same for all galaxies -- irrespective of their morphology --
this gap occurs because elliptical galaxies and the bulges of disc
galaxies inhabit different regions of the $L_{\rm sph} - n$ diagram (see Figure 14 in \citealt{gra2001a}).
That is, for a given light profile shape (i.e. S\'ersic index $n$) the bulges of disc galaxies are brighter than
elliptical galaxies. 
Figure \ref{fig:mbhnpred} allows one to predict that an order of magnitude gap is expected between
the SMBH masses of S\'ersic elliptical galaxies and the S\'ersic bulges of disc galaxies having the same $n$. 

In Figure \ref{fig:mbhnpred} we also show the black hole masses of 51 galaxies that belong to the sample
of \citet{gra2013} but do not have multiple S\'ersic index measurements.
Among them, 13 are core-S\'ersic elliptical galaxies, 5 are core-S\'ersic bulges of disc galaxies,
one is a S\'ersic elliptical galaxy and 32 are S\'ersic bulges of disc galaxies.
We point out that measuring the S\'ersic indices of these galaxies could add many useful
points to the $M_{\rm BH} - n$ diagram. 
In particular, the 13 extra core-S\'ersic elliptical galaxies would allow one 
to better explore the $M_{\rm BH} - n$ diagram in the high-$M_{\rm BH}$ end,
between $10^8$ and $10^{10}~\rm M_\odot$,
where most galaxies are thought to have formed from a different process, namely dry major mergers. 
Similarly, there are an additional 10 S\'ersic bulges of disc galaxies with $M_{\rm BH} < 10^7 ~\rm M_\odot$
that could extend the low-$M_{\rm BH}$ end of the correlation. 

The S\'ersic index is a slight function of the observational bandpass.
This dependency of galaxy structural parameters with 
wavelength arises due to radial gradients in the stellar population gradients and/or dust obscuration \citep{kel2012}.
We therefore plan to perform accurate galaxy image decompositions for all the galaxies 
belonging to the sample of \citet{gra2013} -- with a directly measured SMBH mass -- to explore
the $M_{\rm BH} - n$ relation and other black hole mass scaling relations in a homogeneous analysis (same observational bandpass
and same light profile decomposition method).

Finally, we compare the results from this work with those from \citet{gra2007},
highlighting two main points.
First, and similar to our sample, 
the galaxy sample used by \citet{gra2007} was dominated 
($\sim$80\%) by disc galaxies in the low-mass end ($M_{\rm BH} < 10^8~\rm M_\odot$)
and by elliptical galaxies ($\sim$80\%) in the high-mass end ($M_{\rm BH} > 10^8~\rm M_\odot$).
Second, \citet{gra2007} measured a S\'ersic index greater than 10 for three spheroids with $M_{\rm BH} \sim 10^9~\rm M_\odot$,
which are absent in Figure \ref{fig:mbhn_fit}b.
Combining the different galaxy types and fitting a single relation, it is easy to understand why
a quadratic relation would be more appropriate than a single log-linear relation to describe their data.
At $n=3$ ($M_{\rm BH} \sim 10^8~\rm M_\odot$), their quadratic relation has a slope of $3.70 \pm 0.46$,
similar to that observed for our S\'ersic bulges.

\section{Summary and Conclusions}
\label{sec:concl}
The $M_{\rm BH} - n$ relation \citep{gra2007} is important for
any complete theory or model to describe the coevolution of galaxies and SMBHs.
It also provides a means to estimate black hole masses in galaxies and may prove fruitful    
for recent and future deep, wide-field photometric surveys of galaxies
which can statistically estimate the black hole masses in a large sample of galaxies
up to $z \sim 0.1$.
The main motivation of this work was to re-investigate the $M_{\rm BH} - n$ relation,
given a recent spate of papers which did not detect it.
We have gone beyond the simple recovery of the $M_{\rm BH} - n$ relation, and explored 
potential substructures in this diagram in terms of distinct relations for S\'ersic and
core-S\'ersic galaxies, and for bulges and elliptical galaxies.

We compiled a large collection of literature S\'ersic index measurements 
\citep{gra2007,san2011,vik2012,bei2012} for a sample of 62 galaxies 
with directly measured SMBH masses.
We compared multiple S\'ersic index measurements  
which existed for 35 galaxies,
and found relative 
differences greater than 50\% in many instances.
This is more than expected from a systematic bias produced by different types of light 
profile modelling (1D or 2D) or different observational bandpasses.
We therefore excluded the outlying S\'ersic indices and averaged the remaining values.
This exclusion resulted in the removal of 8 galaxies.
Our final sample therefore consists of 54 galaxies: among them, 27 had S\'ersic indices measured only by one study 
and the remaining 27 have an averaged S\'ersic index measurement. \\

Our principal conclusions are: 
\begin{enumerate}
  \item The $M_{\rm BH} - n$ diagram (Figure \ref{fig:mbhn_fit}) displays an appreciable
  	correlation.
  \item The results from the symmetrical linear regressions (Figure \ref{fig:mbhn_fit}) 
        are consistent at the $2\sigma$ level with predictions (Figure \ref{fig:mbhnpred})
	obtained by combining the $M_{\rm BH} - L_{\rm sph}$ relations for core-S\'ersic 
	and S\'ersic galaxies with the $L_{\rm sph} - n$ relations for elliptical galaxies
	and the bulges of disc galaxies.
  \item If S\'ersic bulges and S\'ersic elliptical galaxies 
  	follow the same $M_{\rm BH} - L_{\rm sph}$ relation, then
  	an order of magnitude gap is expected between
        the SMBH masses of S\'ersic elliptical galaxies 
	and the S\'ersic bulges of disc galaxies having the same $n$. 
\end{enumerate}
A wider range of S\'ersic indices would be beneficial 
to put tighter constraints on the observed slopes of the correlations.
The catalog of 80 directly measured supermassive black hole masses compiled by \citet{gra2013}
allows one to explore the $M_{\rm BH} - n$ diagram in the low- and hig-mass end.
We recognize the need for a well calibrated $M_{\rm BH} - n$ relation
and plan to perform accurate galaxy light profile decompositions
to refine the black hole mass scaling relations. 

\section*{Acknowledgments}
GS thanks Dr. Nicholas Scott for useful discussions.
This research was supported by Australian Research Council funding through grants DP110103509 and FT110100263.
AM, LKH, ES acknowledge support from grants PRIN-MIUR 2010-2011 
``The dark Universe and the cosmic evolution of baryons: from current surveys to Euclid'' 
and PRIN-INAF 2011 ``Black hole growth and AGN feedback through the cosmic time''.
This research has made use of the GOLDMine database \citep{goldmine} and 
the NASA/IPAC Extragalactic Database (NED) which is operated 
by the Jet Propulsion Laboratory, California Institute of Technology, 
under contract with the National Aeronautics and Space Administration.

\clearpage
\onecolumn

\appendix
\section[]{}

Section \ref{sec:aver} illustrates the method we used to combine multiple S\'ersic index measurements 
of the same galaxy. 
These came from four different studies among which only one
(GD07) reported a strong $M_{\rm BH} - n$ relation. 

\begin{figure}
\begin{center}
\includegraphics[width=\columnwidth, trim= 10mm 20mm 10mm 90mm]{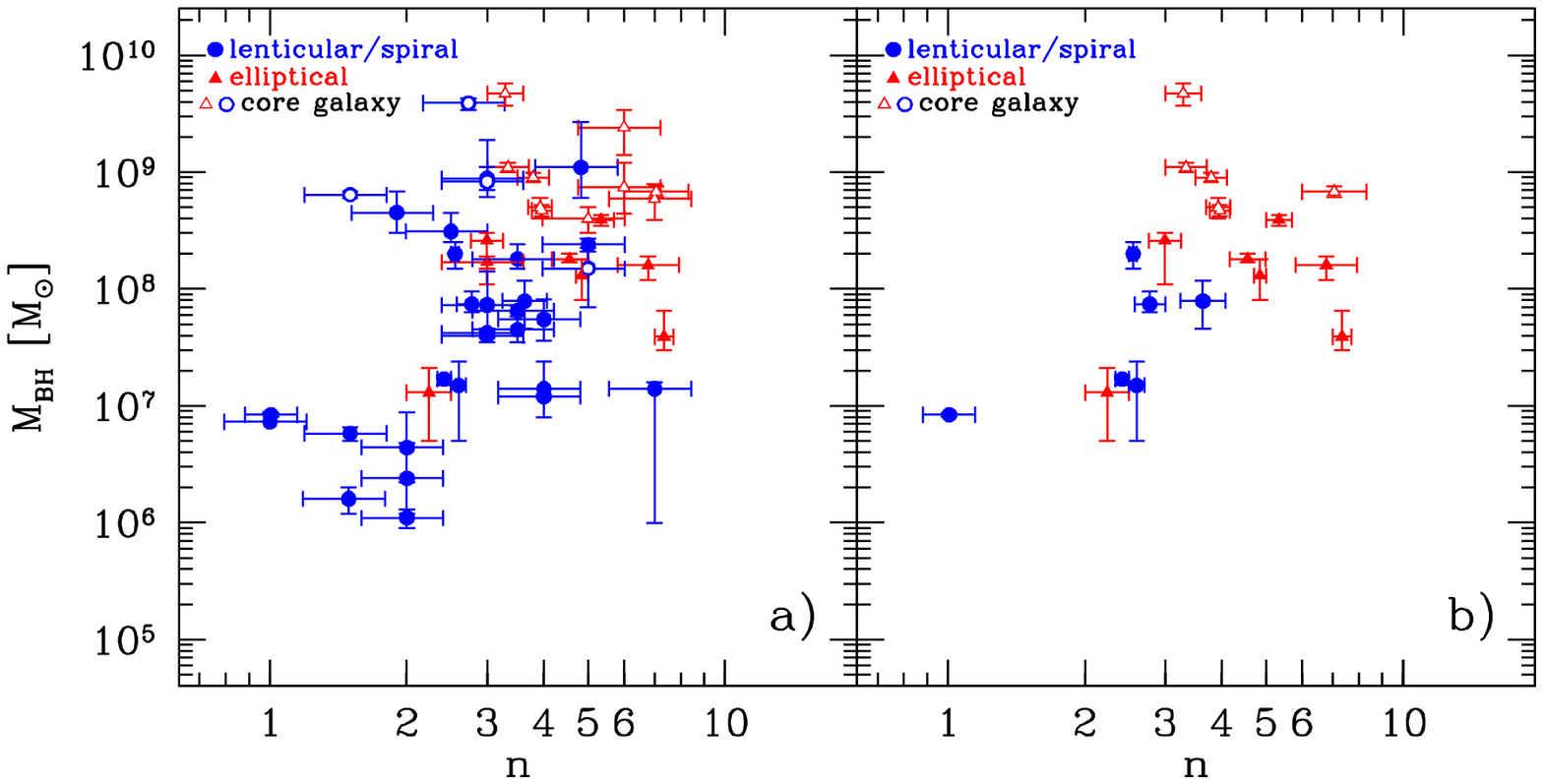}
\end{center}
\caption{Same as Figure \ref{fig:mbhn_fit} but excluding all the GD07 measurements.}
\label{fig:test}
\end{figure}

To check the consistency and the robustness of our results, here we repeat the analysis excluding all the GD07 
measurements. 
Figure \ref{fig:test}a, which can be compared to Figure \ref{fig:mbhn_fit}a, still displays a correlation, although
it is more noisy at the high mass end
(Spearman's correlation coefficient $r_{\rm s}(47) = 0.38$, likelihood of the correlation occuring by chance $P < 1\%$).
Hence we conclude that the inclusion of the GD07 data did not force the recovery of the $M_{\rm BH} - n$ relation(s).
However, the two galaxies previously identified as outliers in Figure \ref{fig:mbhn_fit}b reduce the strength of the 
correlation in Figure \ref{fig:test}b to a likelihood of the correlation occuring by chance to $P < 5\%$.
 
\begin{thebibliography}{99}

\bibitem[Abazajian et al.(2009)]{dr7} Abazajian, K.~N., Adelman-McCarthy, J.~K., Ag{\"u}eros, M.~A., et al.\ 2009, \apjs, 182, 543 
\bibitem[Akritas \& Bershady(1996)]{akr1996} Akritas, M.~G., \& Bershady, M.~A.\ 1996, \apj, 470, 706 
\bibitem[Allen et al.(2006)]{all2006} Allen, P.~D., Driver, S.~P., Graham, A.~W., et al.\ 2006, \mnras, 371, 2
\bibitem[Andredakis et al.(1995)]{and1995} Andredakis, Y.~C., Peletier, R.~F., \& Balcells, M.\ 1995, \mnras, 275, 874 
\bibitem[Beifiori et al.(2012)]{bei2012} Beifiori, A., Courteau, S., Corsini, E.~M., \& Zhu, Y.\ 2012, \mnras, 419, 2497 
\bibitem[Berrier et al.(2013)]{ber2013} Berrier, J.~C., Davis, B.~L., Kennefick, D., et al.\ 2013, \apj, 769, 132
\bibitem[Caon et al.(1993)]{cao1993} Caon, N., Capaccioli, M., \& D'Onofrio, M.\ 1993, \mnras, 265, 1013 
\bibitem[de Vaucouleurs et al.(1991)]{rc3} de Vaucouleurs, G., de Vaucouleurs, A., Corwin, H.~G., Jr., et al.\ 1991, Third Reference 
Catalogue of Bright Galaxies.
\bibitem[Davis et al.(2013)]{dav2013} Davis, B.~L., Berrier, J.~C., Shields, D.~W., Kennefick, J., Kennefick, D., Seigar, M.~S., Lacy, C.~H.~S., \& Puerari, I.\ 2013, \apjs, 199, 33
\bibitem[Driver et al.(2011)]{dri2011} Driver, S.~P., Hill, D.~T., Kelvin, L.~S., et al.\ 2011, \mnras, 413, 971 
\bibitem[Erwin et al.(2004)]{erw2004} Erwin, P., Graham, A.~W., \& Caon, N.\ 2004, Coevolution of Black Holes and Galaxies,  
\bibitem[Ferrarese \& Merritt(2000)]{fer2000} Ferrarese, L., \& Merritt, D.\ 2000, \apjl, 539, L9
\bibitem[Ferrarese et al.(2006a)]{fer2006} Ferrarese, L., C{\^o}t{\'e}, P., Jord{\'a}n, A., et al.\ 2006, \apjs, 164, 334 
\bibitem[Ferrari et al.(2004)]{fer2004} Ferrari, F., Dottori, H., Caon, N., Nobrega, A., \& Pavani, D.~B.\ 2004, \mnras, 347, 824 
\bibitem[Freeman(1970)]{fre1970} Freeman, K.~C.\ 1970, \apj, 160, 811 
\bibitem[Gadotti(2008)]{gad2008} Gadotti, D.~A.\ 2008, \mnras, 384, 420 
\bibitem[Gavazzi et al.(2003)]{goldmine} Gavazzi, G., Boselli, A., Donati, A., Franzetti, P., \& Scodeggio, M.\ 2003, \aap, 400, 451 
\bibitem[Gavazzi et al.(2005)]{gav2005} Gavazzi, G., Donati, A., Cucciati, O., et al.\ 2005, \aap, 430, 411 
\bibitem[Gebhardt et al.(2000)]{geb2000} Gebhardt, K., Bender, R., Bower, G., et al.\ 2000, \apjl, 539, L13 
\bibitem[Graham(2001)]{gra2001a} Graham, A.~W.\ 2001, \aj, 121, 820 
\bibitem[Graham(2008)]{gra2008} Graham, A.~W.\ 2008, \apj, 680, 143 
\bibitem[Graham(2012)]{gra2012} Graham, A.~W.\ 2012, \apj, 746, 113 
\bibitem[Graham(2013)]{gra2013rev} Graham, A.~W. \ 2013, in Planets, Stars and Stellar Systems, Volume 6, p. 91-140, T.~D.~Oswalt \& W.~C.~Keel (Eds.), Springer Publishing (arXiv:1108.0997)
\bibitem[Graham \& Driver(2005)]{gra2005} Graham, A.~W., \& Driver, S.~P.\ 2005, \pasa, 22, 118 
\bibitem[Graham \& Driver(2007a)]{gra2007} Graham, A.~W., \& Driver, S.~P.\ 2007a, \apj, 655, 77
\bibitem[Graham \& Driver(2007b)]{gra2007b} Graham, A.~W., \& Driver, S.~P.\ 2007b, \mnras, 380, L15 
\bibitem[Graham et al.(2007)]{gra2007c} Graham, A.~W., Driver, S.~P., Allen, P.~D., \& Liske, J.\ 2007, \mnras, 378, 198
\bibitem[Graham et al.(2001)]{gra2001b} Graham, A.~W., Erwin, P., Caon, N., \& Trujillo, I.\ 2001, \apjl, 563, L11 
\bibitem[Graham et al.(2003)]{gra2003b} Graham, A.~W., Erwin, P., Trujillo, I., \& Asensio Ramos, A.\ 2003, \aj, 125, 2951 
\bibitem[Graham \& Guzm{\'a}n(2003)]{gra2003a} Graham, A.~W., \& Guzm{\'a}n, R.\ 2003, \aj, 125, 2936 
\bibitem[Graham et al.(2011)]{gra2011b} Graham, A.~W., Onken, C.~A., Athanassoula, E., \& Combes, F.\ 2011, \mnras, 412, 2211 
\bibitem[Graham \& Scott(2013)]{gra2013} Graham, A.~W., \& Scott, N.\ 2013, \apj, 764, 151 
\bibitem[Greene et al.(2008)]{gre2008} Greene, J.~E., Ho, L.~C., \& Barth, A.~J.\ 2008, \apj, 688, 159 
\bibitem[H{\"a}ring \& Rix(2004)]{har2004} H{\"a}ring, N., \& Rix, H.-W.\ 2004, \apjl, 604, L89 
\bibitem[Hunt et al.(2004)]{hun2004} Hunt, L.~K., Pierini, D., \& Giovanardi, C.\ 2004, \aap, 414, 905 
\bibitem[Jerjen et al.(2000)]{jer2000} Jerjen, H., Binggeli, B., \& Freeman, K.~C.\ 2000, \aj, 119, 593 
\bibitem[Kahaner et al.(1989)]{kah1989} Kahaner, D., Moler, C., \& Nash, S. \ 1989, Numerical Methods and Software (Englewood Cliffs: Prentice Hall)
\bibitem[Kelly(2007)]{kel2007} Kelly, B.~C.\ 2007, \apj, 665, 1489 
\bibitem[Kelvin et al.(2012)]{kel2012} Kelvin, L.~S., Driver, S.~P., Robotham, A.~S.~G., et al.\ 2012, \mnras, 421, 1007 
\bibitem[Kormendy \& Richstone(1995)]{kor1995} Kormendy, J., \& Richstone, D.\ 1995, \araa, 33, 581
\bibitem[Laor(2001)]{lao2001} Laor, A.\ 2001, \apj, 553, 677 
\bibitem[Lauer et al.(2007)]{lau2007} Lauer, T.~R., Tremaine, S., Richstone, D., \& Faber, S.~M.\ 2007, \apj, 670, 249 
\bibitem[Lawrence et al.(2007)]{ukirt} Lawrence, A., Warren, S.~J., Almaini, O., et al.\ 2007, \mnras, 379, 1599 
\bibitem[Magorrian et al.(1998)]{mag1998} Magorrian, J., Tremaine, S., Richstone, D., et al.\ 1998, \aj, 115, 2285 
\bibitem[Marconi \& Hunt(2003)]{mar2003} Marconi, A., \& Hunt, L.~K.\ 2003, \apjl, 589, L21
\bibitem[McLure \& Dunlop(2002)]{mcl2002} McLure, R.~J., \& Dunlop, J.~S.\ 2002, \mnras, 331, 795 
\bibitem[M{\'e}ndez-Abreu et al.(2008)]{men2008} M{\'e}ndez-Abreu, J., Aguerri, J.~A.~L., Corsini, E.~M., \& Simonneau, E.\ 2008, \aap, 478, 353 
\bibitem[Pastrav et al.(2013)]{pas2013} Pastrav, B.~A., Popescu, C.~C., Tuffs, R.~J. \& Sansom, A.~E. \ 2013, arXiv:1301.5602
\bibitem[Peng et al.(2002)]{pen2002} Peng, C.~Y., Ho, L.~C., Impey, C.~D., \& Rix, H.-W.\ 2002, \aj, 124, 266 
\bibitem[Peng et al.(2010)]{pen2010} Peng, C.~Y., Ho, L.~C., Impey, C.~D., \& Rix, H.-W.\ 2010, \aj, 139, 2097 
\bibitem[Sani et al.(2011)]{san2011} Sani, E., Marconi, A., Hunt, L.~K., \& Risaliti, G.\ 2011, \mnras, 413, 1479 
\bibitem[S\'ersic(1963)]{ser1963} S\'ersic, J.-L. \ 1963, Boletin de la Asociacion Argentina de Astronomia, 6, 41 
\bibitem[S\'ersic(1968)]{ser1968} S\'ersic, J.-L. \ 1968, Atlas de Galaxias Australes (Cordoba: Observatorio Astronomico)
\bibitem[Scott et al.(2013)]{sgs2013} Scott, N., Graham, A.~W., \& Schombert, J. \ 2013, \apj, 768, 76
\bibitem[Simard et al.(2002)]{sim2002} Simard, L., Willmer, C.~N.~A., Vogt, N.~P., et al.\ 2002, \apjs, 142, 1 
\bibitem[Simard et al.(2011)]{sim2011} Simard, L., Mendel, J.~T., Patton, D.~R., Ellison, S.~L., \& McConnachie, A.~W.\ 2011, \apjs, 196, 11
\bibitem[Trujillo et al.(2004)]{tru2004} Trujillo, I., Erwin, P., Asensio Ramos, A., \& Graham, A.~W.\ 2004, \aj, 127, 1917 
\bibitem[Vika et al.(2012)]{vik2012} Vika, M., Driver, S.~P., Cameron, E., Kelvin, L., \& Robotham, A.\ 2012, \mnras, 419, 2264 
\bibitem[York et al.(2000)]{yor2000} York, D.~G., Adelman, J., Anderson, J.~E., Jr., et al.\ 2000, \aj, 120, 1579
\bibitem[Young \& Currie(1994)]{you1994} Young, C.~K., \& Currie, M.~J.\ 1994, \mnras, 268, L11 

\end{thebibliography}
\end{document}